\documentclass[aps,amsmath,superscriptaddress,citeautoscript,preprint,floatfix]{revtex4}

\usepackage{bm}
\usepackage{amssymb}
\usepackage{graphicx}
\usepackage{amsmath, amsthm, amssymb, graphicx}
\usepackage[usenames]{color}

\newcommand{\LiHoF}{LiHoF$_4$}
\newcommand{\LiHoYF}{LiHo${}_x$Y${}_{1-x}$F${}_4$}

\begin{document}

\title{Direct Observation of Collective Electronuclear Modes About a Quantum Critical Point}

\author{M. Libersky}
\affiliation{Division of Physics, Mathematics, and Astronomy, California Institute of Technology, Pasadena California 91125, USA}
\author{R.D. McKenzie}
\affiliation{Department of Physics and Astronomy, University of British Columbia, Vancouver, British Columbia V6T 1Z1, Canada}
\author{D.M. Silevitch}
\affiliation{Division of Physics, Mathematics, and Astronomy, California Institute of Technology, Pasadena California 91125, USA}
\author{P.C.E. Stamp}
\affiliation{Department of Physics and Astronomy, University of British Columbia, Vancouver, British Columbia V6T 1Z1, Canada}
\affiliation{Pacific Institute of Theoretical Physics, University of British Columbia, Vancouver, British Columbia V6T 1Z1, Canada}
\author{T.F. Rosenbaum}
\email[Correspondence and requests for materials should be addressed to T.F.R.,]{tfr@caltech.edu}
\affiliation{Division of Physics, Mathematics, and Astronomy, California Institute of Technology, Pasadena California 91125, USA}
\date{\today}

\begin{abstract}
    We directly measure the low energy excitation modes of the quantum Ising magnet \LiHoF\ using microwave spectroscopy. Instead of a single electronic mode, we find a set of collective electronuclear modes, in which the spin-$1/2$ Ising electronic spins hybridize with the bath of spin-$7/2$ Ho nuclear spins. The lowest-lying electronuclear mode softens at the approach to the quantum critical point, even in the presence of disorder. This softening is rapidly quenched by a longitudinal magnetic field. Similar electronuclear structures should exist in other spin-based quantum Ising systems.
\end{abstract}

\maketitle


Quantum phase transitions (QPTs) are zero temperature transitions whose critical behavior and fluctuation spectra reveal fundamental properties of technologically useful electronic, magnetic, and optical materials. Canonical examples \cite{hertz76} include the ferromagnet-paramagnet transition in metals, and the quantum Ising model, which describes a set of mutually interacting spin-$1/2$ systems in an `easy axis' crystal field, with quantum fluctuations controlled by an effective field $\Gamma$ perpendicular to the easy axis. Many systems in physics and elsewhere can be mapped to the Ising model in transverse field \cite{jurkevic17,nishimori96,aspuruG20,suzuki13}; recent interest has focused on quantum computing applications \cite{lidar18,bollinger16,lukin17,johnson21}. The model is predicted \cite{hertz76} to have a single spin wave collective mode, whose energy softens to zero exactly at the quantum critical point (QCP).

Although theory predicts that the soft mode must exist, it has never actually been seen near the QCP in any real Ising spin system. One reason for this is defects and paramagnetic impurities, which have a profound effect on QPTs \cite{ji21}. Nuclear spins have a more subtle effect. Many experiments on crystals of transition metal-based magnetic molecules, both in the quantum relaxation regime \cite{villain}, and the high field, low-$T$ regime where spin waves can propagate \cite{taka11}, show that the nuclear spins act as a slowly fluctuating random field \cite{PS98}, which destructively scatters any soft electronic collective mode. 

Rare earth quantum Ising systems have much stronger hyperfine fields, with obvious effects in, e.g., \LiHoYF\ \cite{rosenbaum,giraud,girvin04}. Theory then suggests \cite{moshe05,ryan18} that the pure \LiHoF\ system actually should have 15 coherent electronuclear modes. Instead of scattering the electronic mode, the spin-$7/2$ Ho nuclear spins hybridize with it to create these modes; similar hybridization has been observed in transition-metal antiferromagnets such as CsMnI${}_3$ \cite{Prozorova97}. Nonetheless, previous neutron experiments looking for collective modes in this system \cite{aeppli05} (where there is clear evidence for quantum critical scaling near the QCP \cite{LiHo-QCP}) found only a gapped electronic mode, and no soft mode.


\begin{figure}
    \centering
    \includegraphics[width=240pt]{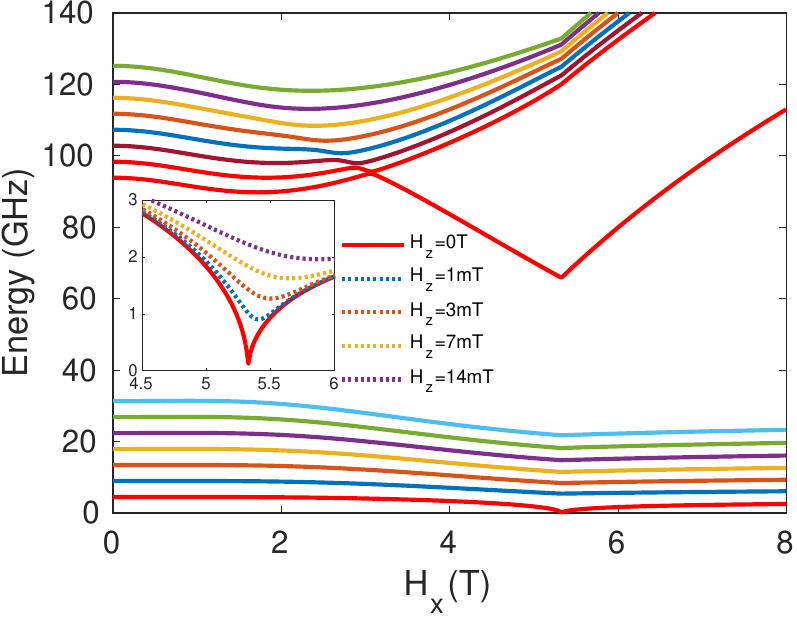}
    \caption{Random Phase Approximation (RPA) calculation of the electronuclear collective mode spectrum at momentum ${\bf k} = 0$ and temperature $T=0$, as a function of transverse field $H_x$, for a long cylinder of \LiHoF. The quantum critical field $H_C \sim 5.3$~T in the calculation. The modes divide into upper and lower groups; at high fields a mode splits off from the upper group. Inset: close-up of the region around the QCP, showing the effect on the soft mode of a small uniform longitudinal field $H_z$.}
    \label{fig:modes}
\end{figure}


The previous theory \cite{moshe05,ryan18} is easily generalized to include the effects of finite $T$ and a small applied longitudinal field $H_z$ \cite{suppInfo}. Salient features, illustrated in Fig. \ref{fig:modes}, include (i) the splitting into upper and lower groups; (ii) the softening of the lowest mode to zero energy when $H_x = H_c$, the transverse field at the QCP; and (iii) the extreme sensitivity of this soft mode to any longitudinal field $H_z$, which immediately gaps the soft mode around the QCP (Fig. \ref{fig:modes}). This last feature has not been discussed previously, and will be of key importance.

Here we describe an experiment on a crystalline sample of \LiHoF, of rectangular prism shape (dimension $1.8 \times 2.5 \times 2.0~\mathrm{mm}^3$), at temperature $T \sim 50$~mK, well below the splitting $\sim 220$~mK between adjacent Ho hyperfine levels. Instead of neutrons, microwave spectroscopy was used, in the frequency range $0.9 < \omega < 5.0$~GHz, to measure AC absorption as a function of $\omega$, $T$, and applied transverse field $H_x$.  A resonator structure is required to amplify the applied ac signal. In order to obtain a high quality factor $Q$ and field homogeneity, we adopted a tunable loop-gap resonator (LGR) design \cite{wood84,libersky19,suppInfo}. The resonant frequencies are tuned by varying the gap capacitance, via partial or complete filling with pieces of sapphire wafer. The incident power level was restricted to $\sim 1~\mu$W (-30 dBm) at the resonator. At this level, sample heating was negligible and the sample was well into the linear response regime.

The spectral weight of the soft mode is predicted to be strongest in the $\chi^{zz}$ configuration \cite{moshe05,ryan18}. This counter-intuitive result is a crystal field effect, and is one reason why the mode was not seen in previous experiments \cite{kovacevic16}. In our setup the resonator and sample are oriented with the AC probe field along the Ising $z$-axis, a solenoid along the transverse $x$-axis, and a split coil along the $z$-axis. In this geometry, crystal fields reduce the AC soft mode absorption along $y$ to zero at the QCP. In the $\chi^{zz}$ configuration the zero mode spectral weight is predicted to diverge  \cite{ryan18} at the critical point when $T=0$; this prediction also holds at the temperatures in our experiment \cite{suppInfo}. However, when one calculates the transmission coefficient $S_{21}(\omega)$ that we measure, this divergence is cancelled by a related divergence in the damping of the magnetopolariton mode formed by the coupling of photons to the soft mode \cite{suppInfo}, where it is also shown that an applied longitudinal field only weakly affects this cancellation.
The cancellation mechanism which leads to the strong suppression of the zero mode in our experiment is reminiscent of cancellation mechanisms in, e.g., the Kondo and spin-boson models\cite{leggett87}; it  also can be related indirectly to the ``light-matter decoupling'' which is hypothesized to exist in cavities \cite{suppInfo,liberato14}.

\begin{figure}
    \centering
    \includegraphics[width=240pt]{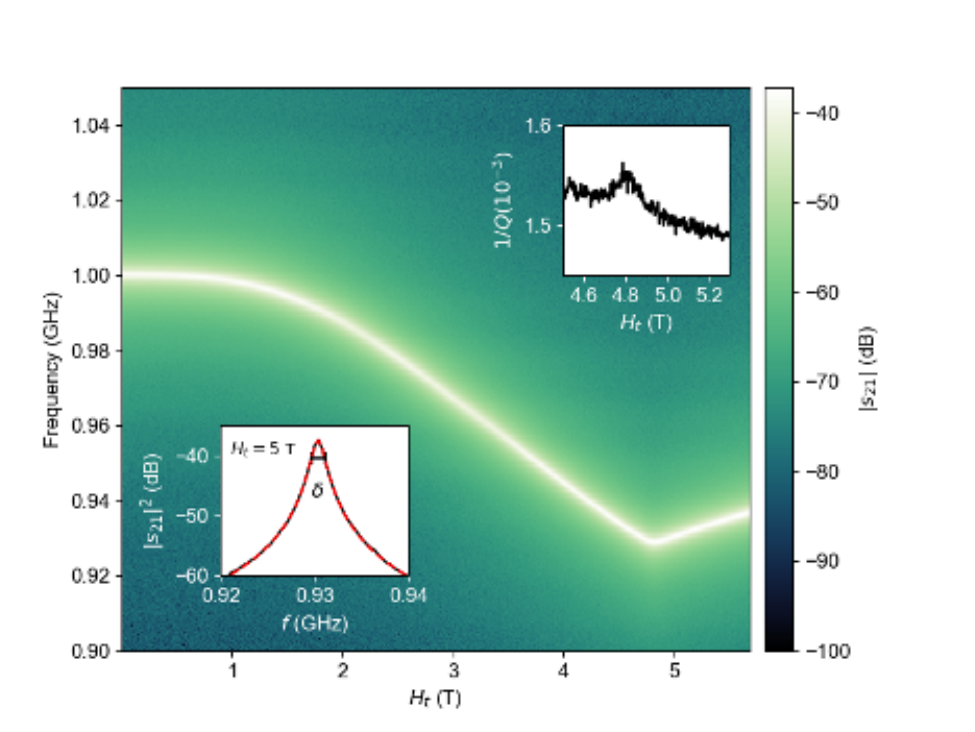}
    \caption{Resonant absorption probing a low-energy excitation mode:  Transmission magnitude $\left|s_{21}\right|^2$ vs. frequency $f$ and transverse magnetic field $H_x$ for a single-mode LGR with zero-field tuning of 1.0 GHz. As the static susceptibility of the \LiHoF\ sample increases with $H_x$, the effective inductance of the resonator + sample circuit increases, resulting in a decreasing resonant frequency, with a cusp at the QPT at $H_C = 4.8$ T. Lower inset: individual frequency spectrum (blue) and Lorentzian fit (orange). Bar indicates the full-width half-maximum point used to determine the quality factor $Q$. Upper inset: $1/Q$ vs. $H_x$, showing enhanced dissipation when the energy of the soft mode matches the 0.93 GHz circuit resonant frequency. }
    \label{fig:lowfreq}
\end{figure}

\begin{figure}
    \centering
    \includegraphics[width=480pt]{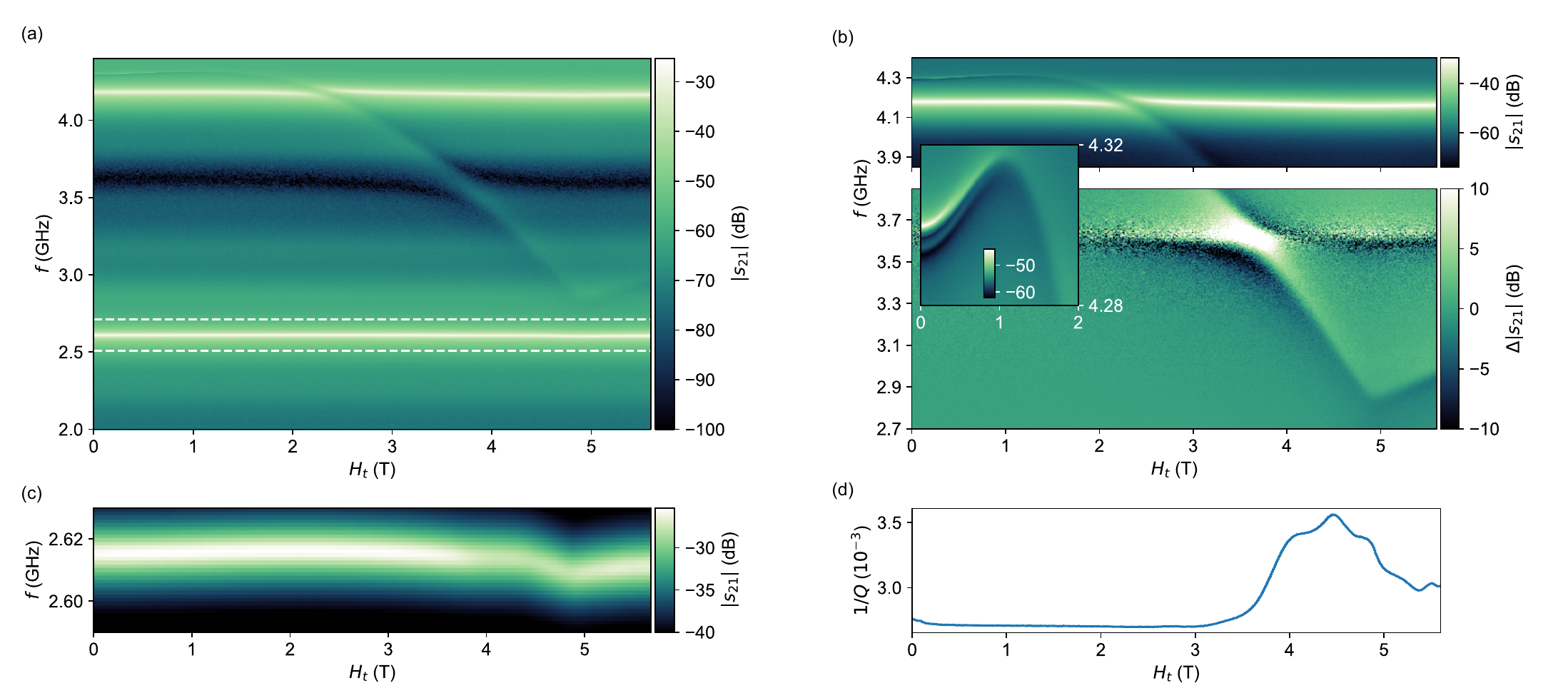}
    \caption{\textbf{Resonant and broadband evolution of higher-energy excitation modes.} (a) Transmission magnitude $\left|s_{21}\right|^2$ vs. frequency and $H_x$, with bimodal resonator tuned to 2.6 and 4.2 GHz. (b) Expanded view of the broadband transmission response. The field evolution of the first excited state response appears as a well-defined continuous curve well away from resonant modes of the LGR. Near the cavity tuning of 4.2 GHz and near an extraneous cavity mode at 3.6 GHz, avoided level crossings can be ascribed to hybridization between cavity photons and magnons. For enhanced contrast, the transmission between 2.7 and 3.8 GHz is plotted relative to a zero-field frequency dependent background $-70$ dB. Inset: Magnified view of transmission in the low-field region where the soft mode and excited states are expected to coincide. A few closely spaced modes are resolved; the non-monotonic shape is reproduced well by the RPA calculations. (c) Expanded view of the resonant response between 2.58 and 2.63 GHz (region between horizontal dashed lines in panel (a)). (d) Transverse field dependence of $1/Q$ for the resonant response shown in (c). At 2.6 GHz, peaks in $1/Q$ are observed above and below the 4.8 T QCP, indicating that at higher frequencies, the soft mode is visible on both sides of the phase transition. For frequencies at and above 2.6 GHz, additional features appear at lower transverse fields as the excited modes intersect with the zero mode.}
    \label{fig:highfreq}
\end{figure}

{\it Results and Analysis}: Figs. \ref{fig:lowfreq}  and  \ref{fig:highfreq} show the  measured  transmission  of  single-crystal  \LiHoF\ in  LGRs tuned to different resonant frequencies at T= 55 mK.  When the resonant frequency $\Omega$ coincides with the soft mode frequency $\omega$, absorption is enhanced, giving a peak in the resonator inverse quality factor $1/Q$  (insets). In Fig. \ref{fig:lowfreq} the resonator is tuned to the lowest accessible $\Omega = 930$ MHz.  In this regime, the field-dependent evolution of the cavity resonant frequency is driven primarily by the change in the static susceptibility of the \LiHoF\ crystal. By varying $\Omega$ we track the soft mode close to the QCP. In Fig. \ref{fig:highfreq} we probe this mode at higher $\Omega$ and find two peaks bracketing the 4.8 T QCP, demonstrating that the mode does persist as expected into the paramagnetic phase.


When $\Omega > 2.8$ GHz, the collective mode-cavity mode coupling is strong enough for detection well away from the cavity resonant frequencies (Fig. \ref{fig:highfreq} (a,b)). We ought to then observe transitions between  all the collective modes, at frequencies equal to their energy differences.  We use a linear combination of absorptive  and dispersive Lorentzian lineshapes, to extract the frequencies and linewidths of these transitions. Near the cavity resonance at 4.2 GHz, the spectra were fit to a coupled oscillator model \cite{schuster10,huebl13}; the apparent avoided level crossing at 3.6 GHz is an anti-resonance in the LGR response.

\begin{figure}
    \centering
    \includegraphics[width=200pt]{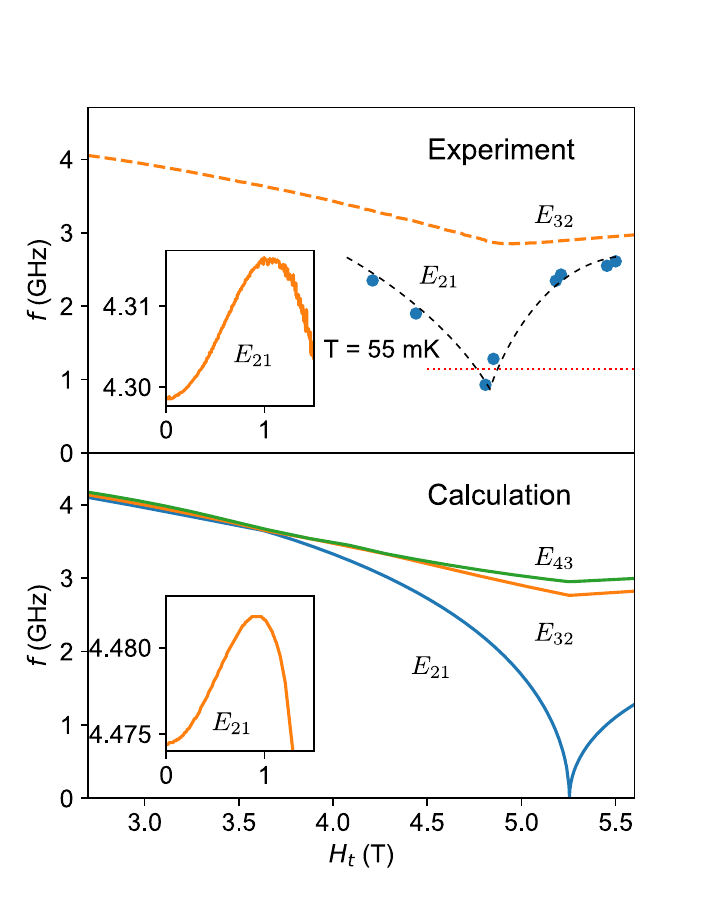}
    \caption{Measured and calculated excitation spectra. Top: Measured field dependence of soft mode ($E_{21}$) and excited state ($E_{32}$) spectra, at $T=55$ mK, as determined by on-resonance (blue points, derived from Fig. \ref{fig:lowfreq} and Fig. \ref{fig:highfreq}(d)) and off-resonance (orange curve, derived from Fig. \ref{fig:highfreq}(b)) responses, respectively. The dashed-line curve through the $E_{21}$ points is a guide to the eye. The horizontal dashed line is the frequency conversion of $T = 55~\mathrm{mK}$.  Bottom: Three lowest transition energies, calculated using a finite-$T$ RPA. The field scale for the QPT differs by $\sim 8\%$. Insets: Measured and calculated frequency evolution at low field, where the three lowest modes are effectively degenerate. The energy scale for the measurement and model differ by $\sim 4\%$. }
    \label{fig:dataoverview}
\end{figure}

We plot in Fig. \ref{fig:dataoverview} the  measured  (top)  and theoretically expected (bottom) transition energies.  The blue points are derived from on-resonance measurements such as those shown in Figs. \ref{fig:lowfreq} and \ref{fig:highfreq}(c,d); the orange curve comes from the broadband measurement shown in Fig. \ref{fig:highfreq}(b).  We note that it is essential to do a finite-$T$ RPA calculation since both the transition energies and their spectral weights differ from their $T = 0$ values. At $T = 55$~mK, which corresponds to  1.15 GHz, one expects multiple transitions between thermally excited electronuclear states \cite{suppInfo}.


At low transverse field, the three lowest excitation modes are essentially degenerate, resulting in a single curve. The insets to Fig. \ref{fig:dataoverview} show this behavior; the non-monotonic field dependence of the measured mode is accurately predicted by the model. The RPA calculations overestimate the critical field, primarily due to the absence of mode-mode couplings in the RPA (which, although individually small, have a cumulative effect on the critical field \cite{ryan18}).

The theoretical result that any applied longitudinal field $H_z$ will gap the soft mode means the domain structure and demagnetization field will play a defining role. In \LiHoF\ the electronic spin dipolar interaction is much larger than the superexchange interaction. One then expects many Ising domains, with thin low-energy domain walls and an almost uniform demagnetization field except very near the boundaries.  This theoretical expectation is confirmed by the observation of micron-sized domains in optical Kerr and Faraday rotation experiments \cite{battison75,pommier88,meyer89}.  The precise structure of the domains \cite{kooy60,gabay85} is then not crucial: what matters is the relation between the mean magnetization density and the demagnetization field. If we model the system as a thick plate, then at zero wavevector, the soft mode is only affected by the average demagnetization field, which we incorporate into the RPA via an effective demagnetizing factor \cite{suppInfo}.

In all the experiments, hysteresis effects were small (in the absence of pinning from impurities, pure \LiHoF\ is a soft ferromagnet). In order to have a well-defined initial state, we defined a magnetic field sweep protocol that always started in the paramagnetic state (with initial $H_x = 5.6$~T, and $H_z = 0$). We then applied a longitudinal field $H_z =70$~mT, lowered $H_x$  to  the  desired  value, and measured the resonator spectra for a series of longitudinal fields.


\begin{figure}
    \centering
    \includegraphics[width=200pt]{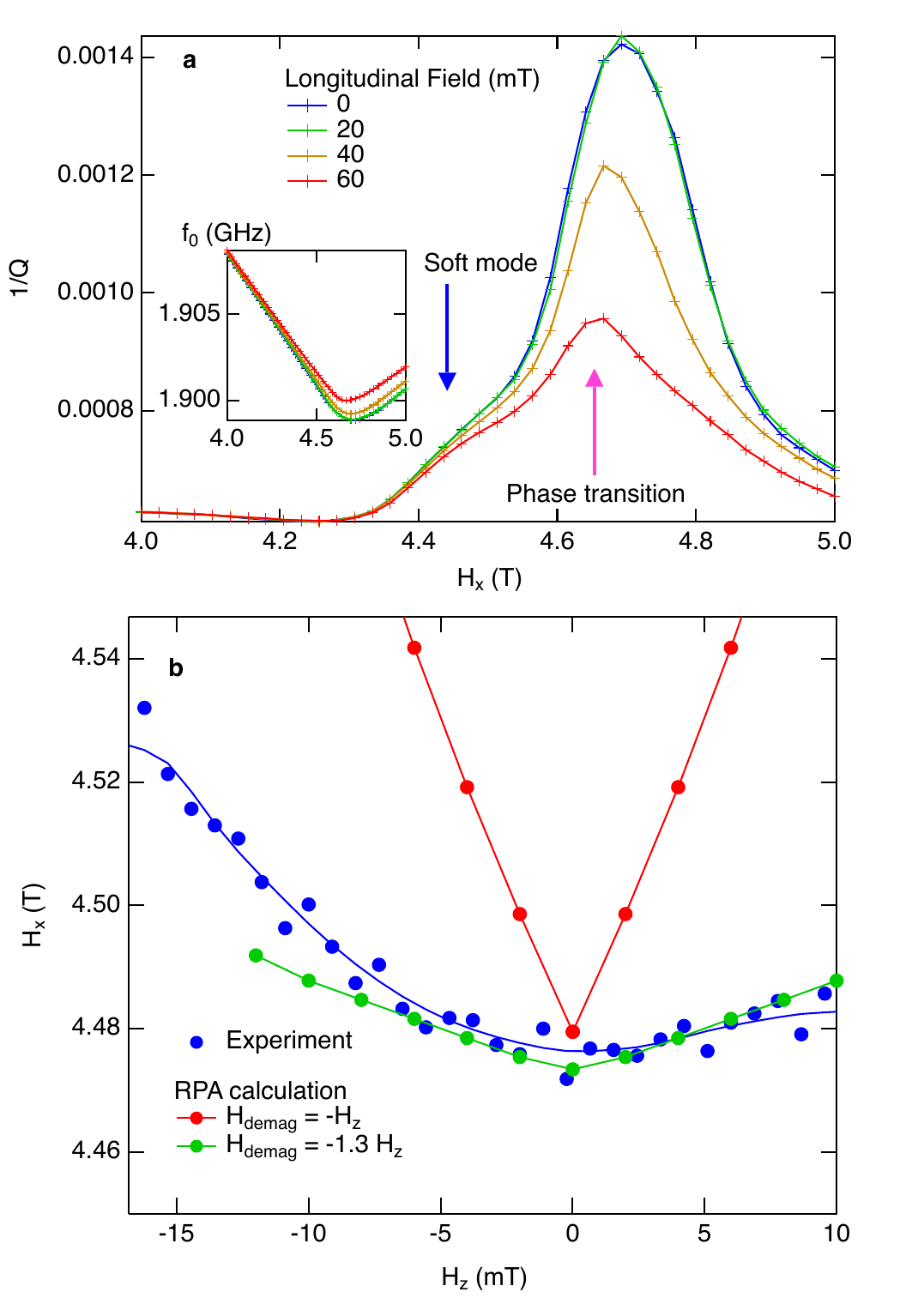}
    \caption{Tuning the soft mode with longitudinal field $H_z$: (a) Dissipation ($1/Q$) near the phase transition and soft mode at 1.9 GHz, following the field-cooling protocol described in the text. The asymmetry in $H_z$ arises from the need to null out geometrical misalignments between the \LiHoF\ crystal and the magnets. The phase transition is marked by the large peak in dissipation at $H_x \sim4.7$ T;  the soft mode appears as a satellite peak at $H_x \sim4.5$ T. Longitudinal (Ising) magnetic fields suppress the dissipation in the main peak, but do not significantly change the amplitude of the soft-mode satellite. Inset: Evolution of resonant frequency for the same set of longitudinal fields. A small shift as a function of $H_z$ is observed. (b) Location of the soft-mode peak for small $H_z$. RPA calculations are for mode locations for two different scalings of internal demagnetization fields. The two theoretical curves are plotted for (i) an internal demagnetization field equal and opposite to the applied field, and (ii) a demagnetization field 30\% larger than the applied field, taking into account a finite domain wall energy. The asymmetry in the measured data is hysteresis due to the field-cooling protocol. At longitudinal fields above 40 mT, domain suppression and the resultant  demagnetization fields lead to non-monotonic behavior (see Fig 3 of Supplementary Information).}
    \label{fig:longfield}
\end{figure}


This protocol is repeated for a series of transverse fields $H_x$ and the resultant mesh of absorptions $1/Q(H_x,H_z)$ is plotted in Fig. \ref{fig:longfield}(a).  We see strong absorption at a critical value of the transverse field for which the lowest energy excitation has a minimum (similar to critical opalescence).  The softening is cut off by $H_z$, substantially suppressing the peak amplitude.  Below the critical value $H_c$ of $H_x$, we also see resonant absorption where the soft mode is degenerate with the cavity mode.  The minimum in the soft mode is then lifted by $H_z$, suppressing its absorption, and reducing the cavity $1/Q$.

Fig. \ref{fig:longfield}(b) compares theory and experiment for the transverse field location of the soft mode minimum at 1.9 GHz. Two theoretical curves are shown. In the first, the average demagnetization field $H_{dm}$ is assumed equal and opposite to $H_z$ (appropriate to zero energy domain walls). This soft mode minimum has a sharper dependence on $H_z$ than seen in experiment. In the second, a finite domain wall energy is assumed. This increases $H_{dm}$ \cite{kooy60}. Micron-sized stripe domains in thin samples of \LiHoF\ indicate a domain wall energy $\sim 10^{-2}~Jm^{-2}$ (which will vary with $H_z, H_x$, and $T$). The actual domain structure will be more complicated (e.g., branching in thick samples \cite{pommier88,gabay85}), but still will increase $H_{dm}$.  Assuming $H_{dm} \sim 1.3 H_z$ (second theoretical curve) yields a good match to the data
in Fig. \ref{fig:longfield}(b).

\vspace{3mm}

{\it Discussion}: The close agreement of theory with experiment indicates that weakly-coupled RPA electronuclear modes represent the true collective degrees of freedom unusually well. Special conditions are required to observe the soft mode: the net longitudinal field $B_z$ in the sample must be homogeneous and zero; we need to measure $\chi^{zz}$; and we need to go to low $\omega,T$. The RPA theory indicates that any net $B_z$ will gap the soft mode.

We can also now identify the gapped mode seen in previous neutron scattering experiments on \LiHoF\ \cite{aeppli05} as the single electronuclear state that splits off from the upper group of modes shown in Fig. \ref{fig:modes}. RPA calculations correctly predict the measured energy of this mode as a function of $H_x$, and also predict it to be the only mode with significant spectral weight at these energies.

There are many systems in which quantum Ising spins couple to both static and dynamic ``defect" modes (spin impurities, two-level systems, nuclear spins, etc.). One example of current interest is in quantum computation. In adiabatic quantum computation the system moves slowly through a QCP \cite{lidar18} such that two-level systems (TLS) are predicted to strongly affect the behavior \cite{amin}. Our results, taken together with previous results on molecular magnet crystals, suggest the following general picture:

(i) When the coupling to these defect modes is weak (as for nuclear spins in transition metal-based molecular magnetic systems like Fe${}_8$, Mn${}_{12}$, V${}_{15}$, etc.), for nuclear spins acting on spin qubits in semiconductors \cite{morello}, or TLS defects weakly coupled to superconductors \cite{simmons04}), then hybridization will be disrupted unless one can go to extremely low $T$. Experiments will then see quantum relaxation of the Ising spins, and no coherent collective modes. To suppress strong decoherence in the Ising spin (qubit) dynamics one must then raise the characteristic qubit operating frequency of these qubits (using, for example, a strong magnetic field \cite{taka11}).

(ii) When the coupling is strong (as for nuclear spins in \LiHoF\ and other rare earth systems, or for some junction TLS defects in superconductors \cite{simmons04}), Ising spin/defect hybridization can occur. If the system is translationally invariant (as in \LiHoF) we then expect coherent hybridized collective modes, one of which will go soft at the QCP. The defects no longer cause decoherence for the Ising spins (qubits), but instead act in concert with them.

Until now there has been no experimental evidence for these coherent modes around a QPT \cite{suhlN}. It remains of considerable interest to investigate and experimentally manipulate them in a variety of magnetic quantum Ising systems. We see that field sweeps through a QPT in adiabatic quantum computing can no longer be regarded as a simple 2 level-avoidance process - one must consider all of the collective modes. Since many such materials are promising candidates for solid-state qubit realizations \cite{bertaina07,morley13,pedersen16}, these collective modes must be characterized fully.

\vspace{3mm}

{\it Acknowledgements}: The experimental work at Caltech was supported by US Department of Energy Basic Energy Sciences Award DE-SC0014866. P.C.E.S. acknowledges support at Caltech from Simons Foundation Award 568762 and National Science Foundation Award PHY-1733907. Theoretical work at UBC was supported by the National Sciences and Engineering Research Council of Canada.


 \end{document}